# Streaking single-electron ionization in open-shell molecules driven by x-ray pulses

M. E. Mountney,[1] T. C. Driver,[2,3] A. Marinelli,[3] M. F. Kling,[2,3,4] J. P. Cryan,[2,3] and A. Emmanouilidou[1]

[1]*Department of Physics and Astronomy, University College London, Gower Street, London WC1E 6BT, United Kingdom*
[2]*Stanford PULSE Institute, SLAC National Accelerator Laboratory, Menlo Park, California 94025, USA*
[3]*SLAC National Accelerator Laboratory, Menlo Park, California 94025, USA*
[4]*Department of Applied Physics, Stanford University, Stanford, California 94305, USA*

We obtain continuum molecular wavefunctions for open-shell molecules in the Hartree-Fock framework. We do so while accounting for the singlet or triplet total spin symmetry of the molecular ion, that is, of the open-shell orbital and the initial orbital where the electron ionizes from. Using these continuum wavefunctions, we obtain the dipole matrix elements for a core electron that ionizes due to single-photon absorption by a linearly polarized x-ray pulse. After ionization from the x-ray pulse, we control or streak the electron dynamics using a circularly polarized infrared (IR) pulse. For a high-intensity IR pulse and photon energies of the x-ray pulse close to the ionization threshold of the $1\sigma$ or $2\sigma$ orbitals, we achieve control of the angle of escape of the ionizing electron by varying the phase delay between the x-ray and IR pulses. For a low-intensity IR pulse, we obtain final electron momenta distributions on the plane of the circularly polarized IR pulse and we find that many features of these distributions correspond to the angular patterns of electron escape solely due to the x-ray pulse.

## I. INTRODUCTION

The development of subfemtosecond extreme ultraviolet (XUV) and x-ray pulses has revolutionized our ability to study electron dynamics in the time domain. For example, using attosecond spectroscopy with co-timed XUV and infrared (IR) pulses it is possible to time-resolve the photoemission process in atomic [1], molecular [2], and solid-state targets [3,4] using the technique known as attosecond streaking [5]. These streaking experiments are typically performed in the low-IR-intensity limit; that is, the ponderomotive energy is lower than the kinetic energy of the electron. However, it has been demonstrated that at higher IR intensities, the properties of the IR field dominate the final momentum distribution and the IR field controls the electron motion [6]. Recently, there has been substantial effort to extend the streaking methodology to x-ray free electron laser (XFEL) sources, which provide a source of high-intensity x-ray pulses, for the purpose of characterizing the temporal structure of the x-ray pulse [7–10]. This includes the development of angular streaking, which employs a circularly polarized IR laser field [11–14]. Similar to studies using table-top sources, XFEL streaking has been extended to study the time-dependent emission pattern of core-excited and core-ionized systems undergoing the Auger-Meitner decay process [14,15], providing insight into electron coherence and entanglement. In this work, we demonstrate the extension of the angular streaking technique [11,14,16,17] to study the time-dependent photoemission in open-shell molecular systems. Most studies of imaging or controlling single-electron ionization instigated by free electron laser (FEL) pulses involve closed-shell molecules [18–21], as they are easier to study compared to open-shell molecules. Indeed, following ionization, for closed-shell molecules, the final molecular ion is in a doublet spin state. For open-shell molecules, the final molecular ion is in a singlet or triplet spin state; that is, the total spin of the open-shell orbital and the orbital where the electron ionizes from is zero or one [22]. In this work, we study single-electron ionization due to single-photon absorption from the open-shell molecule NO when driven by a linearly polarized x-ray pulse. In addition, we streak the electron dynamics with a circularly polarized IR pulse. While previous studies have considered streaking ionization from valence orbitals of NO [23], here we address streaking of single-electron ionization from the core orbitals $1\sigma$ and $2\sigma$. To achieve this, first, in Sec. II A we obtain the dipole matrix element for an electron to ionize following single-photon absorption by an x-ray laser pulse. To compute the dipole matrix element, we describe in Sec. II C how to compute the continuum molecular wavefunctions for open-shell molecules in the Hartree-Fock (HF) framework. We do so while fully accounting for the singlet or triplet symmetry of the final molecular ion. In Sec. II B, we employ the strong-field approximation [24,25] to account for the effect of an IR laser pulse on the dynamics of an electron transitioning from a bound molecular state to the continuum due to an x-ray pulse. Employing the techniques we have developed in Sec. II, using a high-intensity IR pulse we demonstrate control of electron ionization triggered by an x-ray pulse in Sec. III D. That is, we show that there is a one-to-one mapping of the angle of electron escape to the phase delay between the x-ray and IR pulses. We note that a high-intensity (low-intensity) IR pulse refers to an IR pulse with ponderomotive energy that is larger (smaller) than the electron energy following ionization from just the x-ray pulse. Finally, in Sec. III E, we use a low-intensity IR pulse to streak electron ionization by an x-ray pulse. To do so, we obtain doubly differential final electron momenta distributions. We find that several features of these distributions arise from angular patterns of electron ionization solely due to the presence of the x-ray pulse. Such final electron momenta distributions are relevant to the computation of time delays [26,27].

## II. METHOD

### A. Dipole matrix element of an ionizing electron due to single-photon absorption

In the laboratory frame, the dipole matrix element of an electron ionizing due to single-photon absorption is expressed in the length gauge as follows:

$$D_M(\vec{k}') = \sum_{l_1,m_1,m_2,m} e^{i\sigma_{l_1}}(-i)^{l_1} \mathcal{D}^{l_1}_{m_2,m_1}(\hat{R}) \mathcal{D}^{1*}_{M,m}(\hat{R})$$
$$\times Y_{l_1,m_2}(\hat{k}') D_{l_1,m_1,m}, \quad (1)$$

where

$$D_{l_1,m_1,m} = \int d\vec{r}\, \psi^*_{l_1,m_1}(\vec{r};k) \sqrt{\frac{4\pi}{3}} r Y_{1,m}(\hat{r}) \psi_i(\vec{r}). \quad (2)$$

Details on deriving the dipole matrix element are provided in Ref. [6]. In Eqs. (1) and (2), $l_1$ is the angular momentum quantum number, while $m_1$ and $m_2$ are the magnetic quantum numbers. The polarization of the photon in the laboratory and molecular frames are denoted by $M$ and $m$, respectively. Moreover, the vector $\vec{k}' = (k, \theta, \phi)$ is the momentum in the laboratory frame of the electron escaping to the continuum due to single-photon absorption; $\vec{r}$ is the position of this electron in the molecular frame. The energy of the ionizing electron is denoted by $\epsilon$, with $\epsilon = \frac{k^2}{2} = \frac{(k')^2}{2}$. The basis functions $\psi_{l_1 m_1}(\vec{r}; k)$ are the continuum energy molecular eigenstates of a molecule normalized in energy $\epsilon$. Also, $\psi_i(\vec{r})$ are the wavefunctions of the bound orbital $i$ of the molecule under consideration. We express the continuum and bound-state wavefunctions using a single center expansion (SCE) [18,28]

$$\psi(\vec{r}) = \sum_{lm} \frac{P_{lm}(r) Y_{lm}(\theta,\phi)}{r}, \quad (3)$$

where $Y_{lm}$ is a spherical harmonic and $P_{lm}$ is the radial part of the wavefunction. To obtain the continuum wavefunction, we solve a system of HF equations [6,18]. Note that we fully account for the Coulomb potential. The Coulomb phase shift $\sigma_{l_1}(k)$ is given by $\arg \Gamma(l_1 + 1 - \frac{iZ}{k})$, with $Z$ the net charge of the molecular ion resulting after an electron ionizes. In this work we use atomic units unless otherwise stated. In addition, in Eqs. (1) and (2), the matrices $\mathcal{D}$ denote Wigner $D$ functions [29,30], which transform functions from one coordinate system to another. Here, we use the Wigner $D$ functions to rotate from the molecular frame to the laboratory frame. The $z$ axis of the molecular frame is along the principal axis of the molecule. The $z$ axis of the laboratory frame is the same as the polarization direction of the x-ray pulse. In Eq. (1), the Wigner $D$ function $\mathcal{D}^{l_1}_{m_2,m_1}(\hat{R})$ transforms the spherical harmonic of the momentum eigenstate, and $\mathcal{D}^{1*}_{M,m}(\hat{R})$ transforms the spherical harmonic related to the dipole operator. The Euler angles $\hat{R} = (\alpha, \beta, \gamma)$ define the transition from the molecular frame to the laboratory frame. For the Euler angles, we use the convention adapted by Rose [29]. Namely, to transition from the molecular to the laboratory frame, we perform a rotation through an angle $\alpha$ about the $z$ axis, then a rotation through an angle $\beta$ about the new $y$ axis (the $y$ axis after the first rotation), and finally a rotation through an angle $\gamma$ about the new $z$ axis (the $z$ axis after the second rotation). Wigner $D$ functions are the matrix elements of the rotation operator $\mathcal{R} = e^{-i\alpha J_z} e^{-i\beta J_y} e^{-i\gamma J_z}$, i.e.,

$$\begin{aligned}\mathcal{D}^l_{m',m}(\hat{R}) &= \mathcal{D}^l_{m',m}(\alpha,\beta,\gamma) \\ &= \langle lm' | \mathcal{R}(\alpha,\beta,\gamma) | lm \rangle \\ &= e^{-im'\alpha} d^l_{m',m}(\beta) e^{-im\gamma},\end{aligned} \quad (4)$$

where

$$d^l_{m',m}(\beta) = [(l+m')!(l-m')!(l+m)!(l-m)!]^{\frac{1}{2}}$$
$$\times \sum_{s=\max(0,m-m')}^{\min(l+m,l-m')} \left[ \frac{(-1)^{m'-m+s}(\cos\frac{\beta}{2})^{2l+m-m'-2s}}{(l+m-s)!s!} \right.$$
$$\left. \times \frac{(\sin\frac{\beta}{2})^{m'-m+2s}}{(m'-m+s)!(l-m'-s)!} \right]. \quad (5)$$

Details concerning the efficient computation of the function $d^l_{m',m}$ are given in Ref. [31], which we also adopt in this work. For diatomic molecules, the magnetic quantum number is a good number and one can show that Eq. (1) simplifies as follows:

$$D_M(\vec{k}') = \sum_{l_1,m_2,m} e^{i\sigma_{l_1}}(-i)^{l_1} \mathcal{D}^{l_1}_{m_2,m+m_i}(\hat{R}) \mathcal{D}^{1*}_{M,m}(\hat{R})$$
$$\times Y_{l_1,m_2}(\hat{k}') D_{l_1,m+m_i,m}, \quad (6)$$

where $m_i$ is the magnetic quantum number of the bound molecular orbital $i$. The Euler angles are expressed in terms of the polar and azimuthal angles of the symmetry axis of the diatomic molecule as $\hat{R} = (\phi_{\text{mol}}, \theta_{\text{mol}}, 0)$. The total photoionization cross section for a diatomic molecule is given by

$$\sigma_{i \to \epsilon} = \frac{4}{3} \alpha \pi^2 \omega N_i \sum_{M=-1,0,1} \sum_{l_1} D_{l_1,M+m_i,M} D^*_{l_1,M+m_i,M}, \quad (7)$$

where $\alpha$ is the fine structure constant, $N_i$ is the occupation number of the initial molecular orbital $i$, and $\omega$ is the photon energy.

### B. Transition amplitude of an electron from a bound to a continuum state due to combined x-ray and IR pulses

An electron is released into the continuum with momentum $\vec{k}'$ at time $t_{\text{ion}}$ by an x-ray pulse. Then, neglecting the Coulomb potential, the ionizing electron is accelerated by a circular IR laser pulse polarized on the $x$–$z$ plane. Hence, the conserved canonical momentum due to the motion of the electron in the

IR laser field is given by

$$\vec{k}'(t_{\text{ion}}) - \vec{A}_{\text{IR}}(t_{\text{ion}}) = \vec{k}'(t) - \vec{A}_{\text{IR}}(t) = \vec{p}_f, \quad (8)$$

where $\vec{p}_f$ is the final electron momentum at the end of the IR laser field. The vector potential of the IR pulse, $\vec{A}_{\text{IR}}$, is given by

$$\vec{A}_{\text{IR}}(t) = -\frac{E_0^{\text{IR}}}{\omega_{\text{IR}}} \exp\left[-2\log\left(\frac{t}{\tau_{\text{IR}}}\right)^2\right]$$
$$\times \{\sin[\omega_{\text{IR}} t + \phi]\hat{x} + \cos[\omega_{\text{IR}} t + \phi]\hat{z}\}, \quad (9)$$

where $\phi$ is the phase delay between the x-ray and IR pulses, and $E_0^{\text{IR}}$ is the amplitude and $\omega_{\text{IR}}$ the frequency of the electric field of the IR pulse, with $\tau_{\text{IR}}$ being the full width at half maximum (FWHM) in intensity. The envelope of the electric field of the x-ray pulse that ionizes a single electron is given by

$$\vec{E}_X(t) = E_0^X \exp\left[-2\log\left(\frac{t}{\tau_X}\right)^2\right]\hat{z}, \quad (10)$$

where $E_0^X$ is the amplitude and $\tau_X$ is the FWHM in intensity of the x-ray pulse. According to the strong-field approximation (SFA) [24,25], the amplitude for an electron to transition from a bound state $\psi_i$ to a final state with momentum $\vec{p}_f$ in the presence of the x-ray and IR laser fields is given by

$$a(\vec{p}_f) = \int_{t_i}^{t_f} dt_{\text{ion}} E_X(t_{\text{ion}}) D_M(\vec{p}_f + \vec{A}_{\text{IR}}(t_{\text{ion}}))$$
$$\times e^{-iS(t_{\text{ion}}, t_f, \vec{p}_f)}. \quad (11)$$

The times $t_i$ and $t_f$ denote the start and end, respectively, of the IR laser field. The classical action $S$ accumulated during the time interval from $t_{\text{ion}}$ until $t_f$ is given by

$$S(t_{\text{ion}}, t_f, \vec{p}_f) = -I_p t_{\text{ion}} + \int_{t_{\text{ion}}}^{t_f} dt' \frac{[\vec{p}_f + \vec{A}_{\text{IR}}(t')]^2}{2}$$
$$= \frac{p_f^2}{2}(t_f - t_{\text{ion}}) - I_p t_{\text{ion}}$$
$$+ \int_{t_{\text{ion}}}^{t_f} dt \frac{\vec{A}_{\text{IR}}(t) \cdot [\vec{A}_{\text{IR}}(t) + 2\vec{p}_f]}{2}. \quad (12)$$

For details on the computation of $a(\vec{p}_f)$, see Ref. [6]. We compute $a(\vec{p}_f)$ classically using the quantum expression we obtain for $D_M$. We obtain the total amplitude $\mathcal{A}(\vec{p}_f)$ for a certain final electron momentum $\vec{p}_f$ by adding coherently the amplitudes $a_i$ for all trajectories $i$ with the same $\vec{p}_f$ as follows:

$$|\mathcal{A}(\vec{p}_f)|^2 = \left|\sum_i a_i(\vec{p}_f)\right|^2. \quad (13)$$

To compute $\mathcal{A}(\vec{p}_f)$, we first create a two-dimensional grid of the angles of ejection of the electron due to the x-ray pulse in the laboratory frame, $\theta_X$ and $\phi_X$. The polar angle $\theta_X$ ranges from 0° to 180° in steps of 1°, while the azimuthal angle $\phi_X$ ranges from 0° to 360° in steps of 10°. For each $\theta_X$ and $\phi_X$, we produce $2 \times 10^7$ ionization times $t_{\text{ion}}$ using importance sampling [32] in the time interval $[-2.5\tau_X, 2.5\tau_X]$. For the probability distribution in the importance sampling, we use the amplitude of the x-ray pulse at time $t_{\text{ion}}$, i.e., $E_0^X(t_{\text{ion}})$. For each classical trajectory $i$, we propagate the electron in the IR laser field from time $t_{\text{ion}}$ to time $t_f$. Then, we create two final grids. One such grid is in cylindrical coordinates, that is, $(p_{fr}, p_{fy}, \theta)$, and is relevant to our computations in Sec. III D. The momenta components $p_{fy}$ and $p_{fr}$ vary from $-5$ to 5 a.u. and 0 to 5 a.u., respectively, in steps of 0.01 a.u., while the angle $\theta$ varies from 0° to 360° in steps of 1°. The second grid is in Cartesian coordinates, $(p_{fx}, p_{fy}, p_{fz})$, and is relevant to our computations in Sec. III E. All three components of this latter grid vary from $-5$ to 5 a.u. in steps of 0.01 a.u.

### C. Continuum wavefunction for an open-shell molecule

As mentioned above, we obtain the continuum wavefunction by solving a system of HF equations [6,18,33] given by

$$\underbrace{-\frac{1}{2}\nabla^2 \psi_\epsilon(\vec{r}_1)}_{\text{Kinetic energy}} - \underbrace{\sum_n^{\text{nuc}} \frac{Z_n}{|\vec{r}_1 - \vec{R}_n|}\psi_\epsilon(\vec{r}_1)}_{\text{Electron-nuclei}}$$
$$+ \underbrace{\sum_i^{\text{orb}} a_i \int d\vec{r}_2 \frac{\psi_i^*(\vec{r}_2)\psi_i(\vec{r}_2)}{r_{12}}\psi_\epsilon(\vec{r}_1)}_{\text{Direct interaction}}$$
$$- \underbrace{\sum_i^{\text{orb}} b_i \int d\vec{r}_2 \frac{\psi_i^*(\vec{r}_2)\psi_\epsilon(\vec{r}_2)}{r_{12}}\psi_i(\vec{r}_1)}_{\text{Exchange interaction}} = \epsilon \psi_\epsilon(\vec{r}_1), \quad (14)$$

where $r_{12} = |\vec{r}_1 - \vec{r}_2|$, $\psi_\epsilon$ is the continuum wavefunction with energy $\epsilon$ corresponding to a channel $l_1, m_1$, and $\psi_i$ is the bound wavefunction for orbital $i$. The electron-electron interaction involves the direct- and exchange-interaction terms in Eq. (14). Multiplying the terms in Eq. (14) by $\psi_\epsilon^*(\vec{r}_1)$ and integrating over $\vec{r}_1$, we obtain in Dirac notation the following equation:

$$\sum_i^{\text{orb}} a_i J_{i\epsilon} - \sum_i^{\text{orb}} b_i K_{i\epsilon} = \epsilon^{ee} \langle \psi_\epsilon | \psi_\epsilon \rangle, \quad (15)$$

where $\epsilon^{ee}$ is the contribution of the electron-electron interaction terms to the total energy $\epsilon$ and

$$J_{i\epsilon} = \langle \psi_i \psi_\epsilon | \frac{1}{r_{12}} | \psi_i \psi_\epsilon \rangle,$$
$$K_{i\epsilon} = \langle \psi_i \psi_\epsilon | \frac{1}{r_{12}} | \psi_\epsilon \psi_i \rangle, \quad (16)$$

are the direct and exchange terms, respectively. Applying the variational principle [33,34], we obtain

$$\sum_i^{\text{orb}} a_i J_i | \psi_\epsilon \rangle - \sum_i^{\text{orb}} b_i K_i | \psi_\epsilon \rangle = \epsilon^{ee} | \psi_\epsilon \rangle, \quad (17)$$

where

$$J_i | \psi_\epsilon \rangle = \langle \psi_i | \frac{1}{r_{12}} | \psi_i \rangle | \psi_\epsilon \rangle,$$
$$K_i | \psi_\epsilon \rangle = \langle \psi_i | \frac{1}{r_{12}} | \psi_\epsilon \rangle | \psi_i \rangle. \quad (18)$$

The coefficient $a_i$ is the occupation of each bound orbital $i$, with $a_i = 0, 1, 2$ for each molecular orbital. In what follows, we describe how to obtain the coefficients $b_i$.

For closed-shell molecules in the ground state, after an electron ionizes, the resulting molecular ion is left in a doublet state. In the HF framework, the wavefunction of the bound electrons and the ionizing electron can be expressed as a single Slater determinant. However, for open-shell molecules in the ground state, after an electron ionizes from a bound orbital $i$, the resulting molecular ion can be in a singlet or triplet spin state between the open-shell orbital in the ground molecular state and the orbital $i$. Hence, in the HF framework, in order to account for the different spin states of the molecular ion the wavefunction must be written as a linear combination of Slater determinants. Here, we consider the NO open-shell molecule, with electronic configuration ($1\sigma^2$, $2\sigma^2$, $3\sigma^2$, $4\sigma^2$, $1\pi_x^2$, $1\pi_y^2$, $5\sigma^2$, $2\pi^1$). This partially filled $2\pi$ orbital is what leads to singlet and triplet states of the molecular ion after ionization. For a singlet state of the molecular ion between the $i$ and $2\pi$ orbitals, the wavefunction of the molecular ion and the continuum electron is given by the following linear combination of Slater determinants [22]:

$$\Psi(\vec{r}_1, \vec{r}_2, \ldots, \vec{r}_N) = \frac{1}{\sqrt{2}}[|\{\psi_j\}_{j \neq i, 2\pi} \psi_i^\uparrow \psi_{2\pi}^\downarrow \psi_\epsilon^\uparrow| \\ - |\{\psi_j\}_{j \neq i, 2\pi} \psi_i^\downarrow \psi_{2\pi}^\uparrow \psi_\epsilon^\uparrow|], \quad (19)$$

where $\vec{r}_1, \vec{r}_2, \ldots, \vec{r}_N$ are the positions of the electrons and $\{\psi_j\}_{j \neq i, 2\pi}$ are the 12 spin orbital wavefunctions of the electrons, not including the orbitals $i$ and $2\pi$. Orbitals with spin $\frac{1}{2}$ are denoted by an up arrow and orbitals with spin $-\frac{1}{2}$ are denoted by a down arrow. Next, we verify that the wavefunction in Eq. (19) indeed corresponds to a singlet state by applying the spin projection operator $\hat{S}_z$ and the total spin operator $\hat{S}^2$. As shown in Ref. [35], application of the $\hat{S}_z$ operator to a Slater determinant $\Phi$ results in

$$\hat{S}_z \Phi = M_S \Phi, \\ M_S = \tfrac{1}{2}(n_\alpha - n_\beta), \quad (20)$$

where $n_\alpha$ and $n_\beta$ are the number of columns with spin up and spin down, respectively. Moreover, as shown in Ref. [35], applying the operator $\hat{S}^2$ on $\Phi$ results in

$$\hat{S}^2 \Phi = \left\{ \sum_P \hat{P}_{\alpha\beta} + \frac{1}{4}[(n_\alpha - n_\beta)^2 + 2n_\alpha + 2n_\beta] \right\} \Phi, \quad (21)$$

where $\hat{P}_{\alpha\beta}$ is an operator interchanging opposite $\alpha$ and $\beta$ spins in $\Phi$. One can show using the properties of determinants that applying these spin operators to the states $\{\psi_j\}_{j \neq i, 2\pi}$ equates to zero, meaning it is sufficient to apply the operators only to the three-particle system between the orbitals $i$, $2\pi$, and $\epsilon$. Hence, the eigenvalue for the $\hat{S}_z$ on the state $\Psi$ is given by

$$\hat{S}_z \Psi = \frac{1}{\sqrt{2}}(\hat{S}_z |\psi_i^\uparrow \psi_{2\pi}^\downarrow \psi_\epsilon^\uparrow| - \hat{S}_z |\psi_i^\downarrow \psi_{2\pi}^\uparrow \psi_\epsilon^\uparrow|) \\ = \frac{1}{\sqrt{2}}\left(\frac{1}{2}(2-1)|\psi_i^\uparrow \psi_{2\pi}^\downarrow \psi_\epsilon^\uparrow| - \frac{1}{2}(2-1)|\psi_i^\downarrow \psi_{2\pi}^\uparrow \psi_\epsilon^\uparrow|\right) \\ = \frac{1}{2}\Psi. \quad (22)$$

This is consistent with the spin of the $2\pi$ and $i$ orbitals being zero and the spin of the continuum electron being $+\frac{1}{2}$. Next, the eigenvalue of the $\hat{S}^2$ operator when applied to the $\Psi$ singlet state is given by

$$\hat{S}^2 \Psi = \frac{1}{\sqrt{2}}(\hat{S}^2 |\psi_i^\uparrow \psi_{2\pi}^\downarrow \psi_\epsilon^\uparrow| - \hat{S}^2 |\psi_i^\downarrow \psi_{2\pi}^\uparrow \psi_\epsilon^\uparrow|) \\ = \frac{1}{\sqrt{2}}\bigg[|\psi_i^\downarrow \psi_{2\pi}^\uparrow \psi_\epsilon^\uparrow| + |\psi_i^\uparrow \psi_{2\pi}^\uparrow \psi_\epsilon^\downarrow| \\ + \frac{1}{4}((2-1)^2 + 4 + 2)|\psi_i^\uparrow \psi_{2\pi}^\downarrow \psi_\epsilon^\uparrow|\bigg] \\ - \frac{1}{\sqrt{2}}\bigg[|\psi_i^\uparrow \psi_{2\pi}^\downarrow \psi_\epsilon^\uparrow| + |\psi_i^\uparrow \psi_{2\pi}^\uparrow \psi_\epsilon^\downarrow| \\ + \frac{1}{4}((2-1)^2 + 4 + 2)|\psi_i^\downarrow \psi_{2\pi}^\uparrow \psi_\epsilon^\uparrow|\bigg] \\ = \frac{3}{4}\Psi. \quad (23)$$

This is consistent with the $i$ and $2\pi$ orbitals having spin equal to zero and the continuum electron having total spin $+\frac{1}{2}$. Using the wavefunction $\Psi$, one can show that obtaining the $b_i$ coefficients is equivalent to obtaining the $b_i$ coefficients from two limiting cases. The first case involves the electrons in bound orbitals $i$ and $2\pi$ and the continuum electron $\epsilon$, while the other case involves the two electrons in closed orbitals $j$ and the continuum electron. For the first case, we consider the three-electron wavefunction

$$\Psi(\vec{r}_1, \vec{r}_2, \vec{r}_3) = \frac{1}{\sqrt{2}}\left( \frac{1}{\sqrt{3!}} \begin{vmatrix} \psi_i^\uparrow(\vec{r}_1) & \psi_{2\pi}^\downarrow(\vec{r}_1) & \psi_\epsilon^\uparrow(\vec{r}_1) \\ \psi_i^\uparrow(\vec{r}_2) & \psi_{2\pi}^\downarrow(\vec{r}_2) & \psi_\epsilon^\uparrow(\vec{r}_2) \\ \psi_i^\uparrow(\vec{r}_3) & \psi_{2\pi}^\downarrow(\vec{r}_3) & \psi_\epsilon^\uparrow(\vec{r}_3) \end{vmatrix} \\ - \frac{1}{\sqrt{3!}} \begin{vmatrix} \psi_i^\downarrow(\vec{r}_1) & \psi_{2\pi}^\uparrow(\vec{r}_1) & \psi_\epsilon^\uparrow(\vec{r}_1) \\ \psi_i^\downarrow(\vec{r}_2) & \psi_{2\pi}^\uparrow(\vec{r}_2) & \psi_\epsilon^\uparrow(\vec{r}_2) \\ \psi_i^\downarrow(\vec{r}_3) & \psi_{2\pi}^\uparrow(\vec{r}_3) & \psi_\epsilon^\uparrow(\vec{r}_3) \end{vmatrix} \right). \quad (24)$$

To obtain the electron-electron interaction terms in Eqs. (14) and (15), we compute

$$\langle \Psi | \frac{1}{r_{12}} + \frac{1}{r_{13}} + \frac{1}{r_{23}} | \Psi \rangle, \quad (25)$$

which we find to be equal to

$$\langle \Psi | \frac{1}{r_{12}} + \frac{1}{r_{13}} + \frac{1}{r_{23}} | \Psi \rangle = \frac{1}{2}(2J_{i,2\pi} + 2J_{i,\epsilon} + 2J_{2\pi,\epsilon} - K_{i,\epsilon} \\ - K_{2\pi,\epsilon} + 2K_{i,2\pi}). \quad (26)$$

Using the variational principle with respect to $\psi_\epsilon$, we obtain

$$\left(J_i - \tfrac{1}{2}K_i + J_{2\pi} - \tfrac{1}{2}K_{2\pi}\right)\psi_\epsilon = \epsilon^{ee}\psi_\epsilon. \quad (27)$$

Comparing Eq. (17) with Eq. (27), we find $a_i = 1$, $a_{2\pi} = 1$, $b_i = \frac{1}{2}$, and $b_{2\pi} = \frac{1}{2}$. The other limiting case involves two electrons in a closed shell $j$ and a continuum electron. In this

TABLE I. Ionization energies of the $1\sigma$, $2\sigma$, and $4\sigma$ orbitals for NO.

| Orbital | This work (eV) | Experiment (eV) |
|---|---|---|
| $1\sigma$ $^1\Pi$ | 543.5 | 543.8 [4] |
| $1\sigma$ $^3\Pi$ | 543.1 | 543.3 [4] |
| $2\sigma$ $^1\Pi$ | 411.6 | 411.8 [4] |
| $2\sigma$ $^3\Pi$ | 410.2 | 410.3 [4] |
| $4\sigma$ $^1\Pi$ | 23.0 | 21.8 [40] |
| $4\sigma$ $^3\Pi$ | 21.2 | 21.7 [40] |

case, the three-electron wavefunction is given by

$$\Psi(\vec{r}_1,\vec{r}_2,\vec{r}_3) = \frac{1}{\sqrt{3!}} \begin{vmatrix} \psi_j^{\uparrow}(\vec{r}_1) & \psi_j^{\downarrow}(\vec{r}_1) & \psi_{\epsilon}^{\uparrow}(\vec{r}_1) \\ \psi_j^{\uparrow}(\vec{r}_2) & \psi_j^{\downarrow}(\vec{r}_2) & \psi_{\epsilon}^{\uparrow}(\vec{r}_2) \\ \psi_j^{\uparrow}(\vec{r}_3) & \psi_j^{\downarrow}(\vec{r}_3) & \psi_{\epsilon}^{\uparrow}(\vec{r}_3) \end{vmatrix}. \quad (28)$$

Using the same method as above, we find $a_j = 2$ and $b_j = 1$. Finally, one can show that for the triplet state of the molecular ion between orbitals $i$ and $2\pi$, the wavefunction is given by

$$\Psi = \frac{1}{\sqrt{6}}[2|\{\psi_j\}_{j\neq i,2\pi}\psi_i^{\uparrow}\psi_{2\pi}^{\uparrow}\psi_{\epsilon}^{\downarrow}| - |\{\psi_j\}_{j\neq i,2\pi}\psi_i^{\downarrow}\psi_{2\pi}^{\uparrow}\psi_{\epsilon}^{\uparrow}| \\ - |\{\psi_j\}_{j\neq i,2\pi}\psi_i^{\uparrow}\psi_{2\pi}^{\downarrow}\psi_{\epsilon}^{\uparrow}|]. \quad (29)$$

Following the same method as above, we find that $a_i = 1$, $b_i = -\frac{1}{2}$, $a_{2\pi} = 1$, $b_{2\pi} = -\frac{1}{2}$, $a_j = 2$, and $b_j = 1$.

## III. RESULTS

### A. Computation of the bound and continuum orbitals

First, we compute the ionization energies of the $1\sigma$, $2\sigma$, and $4\sigma$ orbitals. We do so using the complete active space self-consistent field (CASSCF) method [36] in the framework of the quantum chemistry package MOLPRO [37]. In MOLPRO, we employ the augmented Dunning correlation consistent quadruple valence basis set (aug-cc-pVQZ) [38]. We find the equilibrium distance of the ground state of NO to be equal to 1.1508 Å, in agreement with Ref. [39]. In Table I, we compare the ionization energies we obtain with experimental results [4,40]. Table I shows that our results agree very well with experimental ones, particularly for the $1\sigma$ and $2\sigma$ orbitals. We consider single-photon absorption due to the x-ray pulse from the core orbitals $1\sigma$ and $2\sigma$, which are localized on the oxygen and nitrogen sides, respectively. We also obtain the single-photon ionization cross section for the valence orbital $4\sigma$. For the calculation of the photoionization cross section from the $4\sigma$ orbital, we use the bound orbitals of the ground state of NO. However, for ionization from the $1\sigma$ and $2\sigma$ orbitals, we use the bound orbitals of the excited state of $NO^+$ with a core hole in the $1\sigma$ or $2\sigma$ orbital. The reason is that for ionization from the $1\sigma$ or $2\sigma$ orbital the hole state is highly localized. This causes the electron density of the remaining electrons to adjust to this new potential. For ionization from the $4\sigma$ orbital, the hole is delocalized, so the effect of orbital relaxation is less noticeable (see Refs. [4,41,42]). We compute the bound states using the HF method with the aug-cc-pVQZ basis set. Moreover, we find that in the single center expansion [see

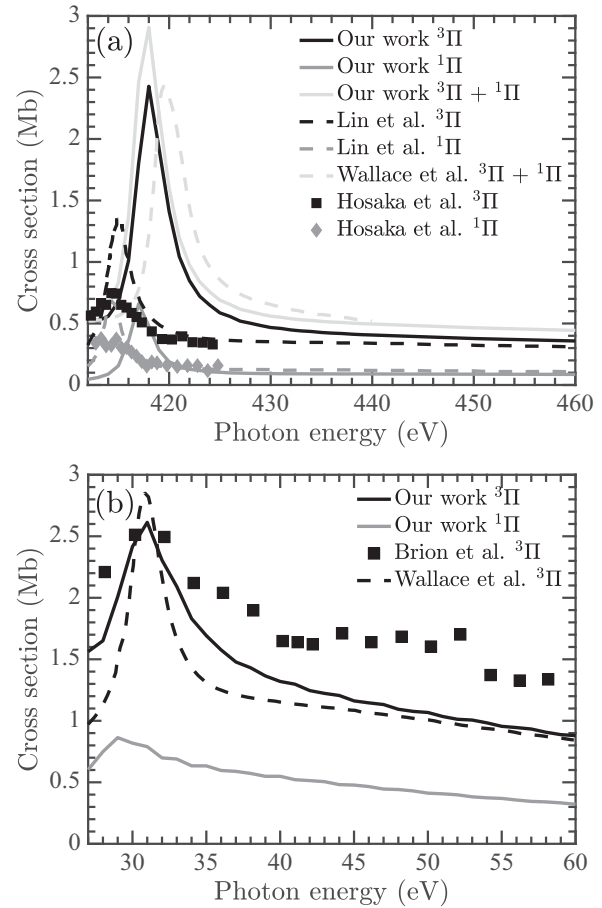

FIG. 1. Photoionization cross sections for NO (a) from the $2\sigma$ orbital and (b) from the $4\sigma$ orbital. In (a), we compare the results we obtain for the singlet state of $NO^+$ (solid dark grey line) and for the triplet state of $NO^+$ (solid black line) with the theoretical results of Lin and Lucchese [41] for the singlet (dashed dark grey line) and triplet (dashed black line) of $NO^+$, as well as with the experimental results of Hosaka *et al.* [43] (dark grey diamonds and black squares). The total cross section we obtain for the singlet plus the triplet state of $NO^+$ (solid light grey line) is compared with the theoretical results of Wallace *et al.* [45] (dashed light grey line). In (b), we compare the results we obtain for the triplet state of $NO^+$ (solid black line) with the theoretical results of Wallace *et al.* [45] (dashed black line), as well as with the experimental results of Brion and Tan [44] (black squares). We also show the results we obtain for the singlet state of $NO^+$ (solid dark grey line).

Eq. (3)] it suffices to truncate the expansion over the $l$ quantum numbers up to $l_{\max} = 30$ for the bound wavefunctions and $l_{\max} = 19$ for the continuum wavefunctions. This truncation ensures convergence of the single-photon ionization cross sections. Also, we note that in the computation of the dipole matrix element in Eq. (6) for the $1\sigma$, $2\sigma$, and $4\sigma$ states, we use $m_i = 0$.

### B. Photoionization cross sections

In Fig. 1, we compare the total photoionization cross sections from the $2\sigma$ and $4\sigma$ orbitals obtained in this work using Eq. (7) with the experimental results of Refs. [43,44], as well as with the theoretical results of Refs. [41,45]. For ionization from the $2\sigma$ orbital, the cross sections we obtain for the singlet and triplet states [Fig. 1(a)] exhibit a shape resonance [46] at roughly 417 and 418 eV, respectively, which are a few eV higher than the resonances obtained in the theoretical work of Lin and Lucchese [41]. We find that besides this offset in the location of the shape resonance, the overall shape of the cross section for the singlet state with respect to the photon energy is in very good agreement with the cross section obtained by Lin and Lucchese [41]. However, the maximum cross section we obtain for the triplet state is roughly two times higher than the one obtained by Lin and Lucchese [41]. For large photon energies, the results we obtain for the singlet and triplet states agree with Lin and Lucchese [41] as well as with the experimental results [43]. The total cross section of the singlet plus the triplet states of $NO^+$ obtained in this work is found to be in close agreement with the theoretical result of Wallace *et al.* [45]. We note that the theoretical technique we use to obtain the cross sections is more accurate than the method employed by Wallace *et al.* [45] but less accurate than the one used by Lin and Lucchese [41]. Specifically, the work of Wallace *et al.* [45] employs the multiple scattering method (MSM), where the cross section of the singlet and triplet states differs only by the spin-statistical ratio of 1:3. Lin and Lucchese [41] compute the continuum wavefunctions separately for the singlet and triplet states using the multichannel Schwinger configuration interaction (MCSCI) method. In our work, we compute the continuum wavefunctions in the Hartree-Fock framework using different coefficients $b_i$ for the singlet and triplet states, as discussed in Sec. II C. Finally, we multiply by the spin-statistical ratio 1:3 the cross sections of the singlet and triplet states of $NO^+$. In Fig. 1(b), we compare our results for the photoionization cross section from the $4\sigma$ orbital for the triplet state of $NO^+$ with the theoretical result of Wallace *et al.* [45] and with the experimental result of Brion and Tan [44]. We find that all results exhibit a shape resonance at roughly 31 eV and have similar values for high photon energies. However, the cross section we obtain for the triplet state of $NO^+$ has a better agreement with the experiment [44] compared to the cross section obtained by Wallace *et al.* [45].

### C. Photoionization by the x-ray pulse

In what follows, we obtain the differential cross section for an electron to ionize by single-photon absorption from the O side, $1\sigma$ orbital, or the N side, $2\sigma$ orbital, only due to the x-ray laser field. In Fig. 2, we plot the absolute value square of $D_M(\vec{k}')$ in Eq. (6), which is proportional to the differential cross section. Since we consider a linearly polarized x-ray pulse, the polarization of the photon in the laboratory frame $M$ is equal to zero. We do so for two different photon energies: one close to threshold, i.e., 546 eV for the $1\sigma$ and 413 eV for the $2\sigma$ orbital, and for a photon energy significantly higher than the ionization threshold, i.e., 623 eV for the $1\sigma$ and 490 eV for the $2\sigma$ orbital. As expected, we find that for the high photon energies the electron has significantly higher probability to ionize towards the side where the electron originally ionizes from. That is, for ionization from the $1\sigma$ orbital the electron mostly escapes towards the O side for a photon energy of 623 eV. For ionization from the $2\sigma$ orbital, the electron mostly ionizes along the N side for a photon energy of 490 eV. However, for lower photon energies the electron has significant probability to ionize towards the other side as well [see Figs. 2(a) and 2(c)].

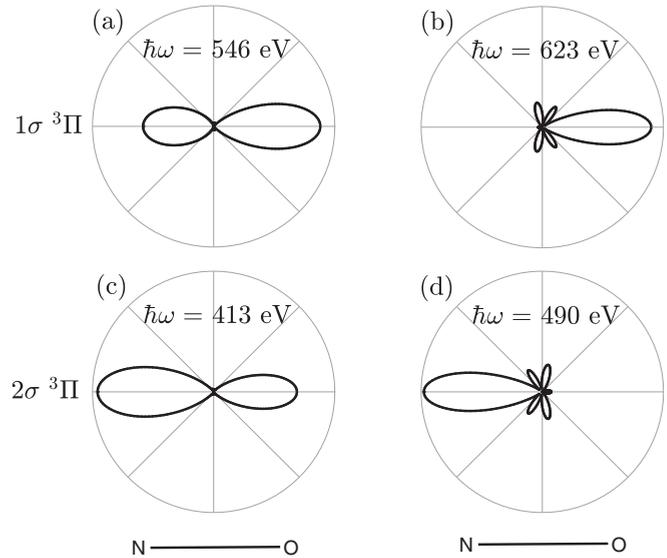

FIG. 2. Differential cross section for an electron to ionize at a certain angle $\theta_X$ from [(a), (b)] the $1\sigma$ orbital and [(c), (d)] the $2\sigma$ orbital for the triplet state of $NO^+$.

### D. Control of electron ionization triggered by an x-ray pulse

Coherent control is an important tool with wide applications in quantum optics and metrology [47–49], attosecond metrology [50,51], optoelectronics [52], and laser cooling [53,54]. Recent studies [55–58] succeeded in steering the direction of electron current by controlling the quantum interference of excitation or ionization pathways resulting from a mid-IR $\omega$ pulse and its second harmonic $2\omega$ [55]. Control of the interference between the two-photon ($\omega$) and single-photon ($2\omega$) pathways and finally of the direction of electron motion was achieved using as a tool the phase delay of the two pulses [55]. In this optical technique, the dimensions over which the electron current is generated are limited to roughly one wavelength of the infrared light that is used to accelerate the electrons [59], i.e., to a few micrometers. Using coherent control of one- and two-photon processes to reduce to the nanometer scale the dimensions over which current is produced requires optically generating $\omega$ and $2\omega$ vacuum ultraviolet (VUV) light beams. This is currently impractical. In what follows, we theoretically demonstrate that control of electron currents generated at roughly a few nanometers is possible. We demonstrate control by varying the phase delay between a linearly polarized x-ray pulse and a circularly polarized IR pulse, while keeping the orientation of the NO molecule fixed and parallel to the linear polarization of the x-ray pulse. Specifically, for high intensities of the IR pulse we find a one-to-one correspondence between the final angle of electron escape and the phase delay between the x-ray and IR pulses. This was shown in Ref. [6] in the context of $N_2$.

We obtain the probability for an electron to escape to the continuum on the $x-z$ plane of the IR pulse with an angle

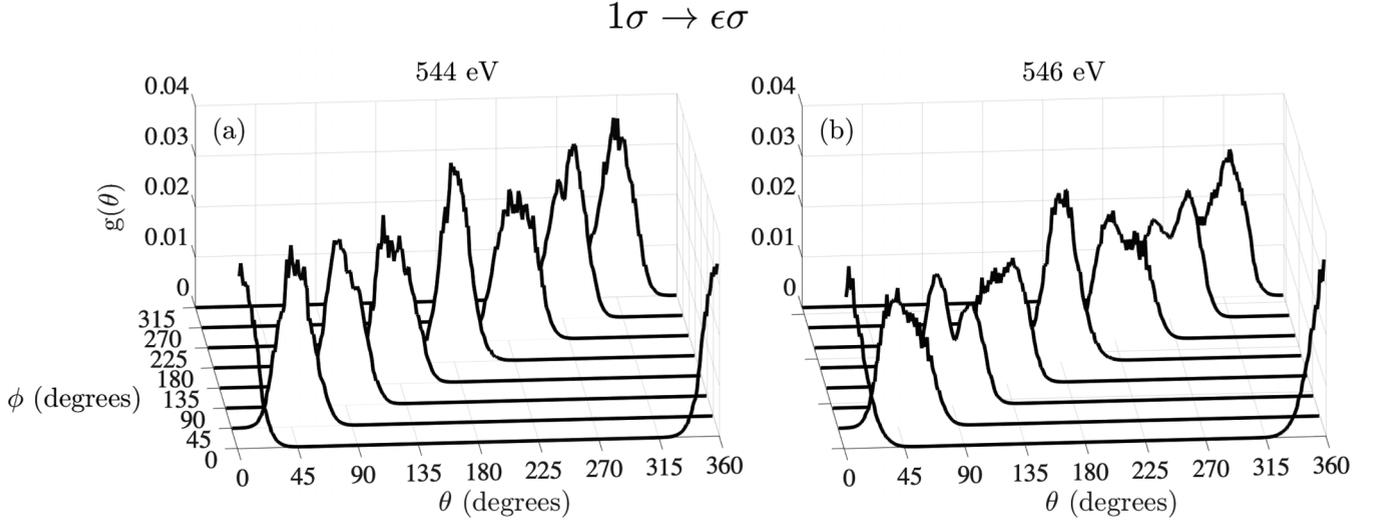

FIG. 3. For the triplet state of NO$^+$, we waterfall plot the probability $g(\theta)$ as a function of the angle $\theta$ of electron escape for different phase delays $\phi$ between the x-ray and IR pulses. The electron ionizes from the $1\sigma$ orbital of NO and the intensity of the circularly polarized IR pulse is $5 \times 10^{13}$ W/cm$^2$. The photon energy of the x-ray pulse is (a) 544 eV and (b) 546 eV.

$\theta$. This angle is measured with respect to the $z$ axis in the laboratory frame. In this section, the NO molecule lies along the $z$ axis. To obtain the probability for an electron to ionize on the $x-z$ plane with momentum $(p_{fr}, \theta)$, we integrate $|\mathcal{A}(\vec{p}_f)|^2$ from Eq. (13) over the $p_{f_y}$ component,

$$|\mathcal{A}(p_{fr}, \theta)|^2 = \int dp_{fy} |\mathcal{A}(\vec{p}_f)|^2. \quad (30)$$

Then, integrating $|\mathcal{A}(p_{fr}, \theta)|^2$ in Eq. (30) over $p_{fr}$, we find that $g(\theta)$, the probability for an electron to ionize with angle $\theta$ on the $x-z$ plane, is given as follows:

$$g(\theta) = \int dp_{f_r} p_{f_r} |\mathcal{A}(p_{fr}, \theta)|^2. \quad (31)$$

In Figs. 3 and 4, we plot the probability $g(\theta)$ for a high intensity of the circularly polarized IR pulse, equal to $5 \times 10^{13}$ W/cm$^2$. Here, the molecular axis of the NO molecule is parallel to the x-ray pulse. We consider a photon energy of the IR pulse $\omega_{IR}$ equal to 2300 nm, while the FWHM is $\tau_{IR} = 100$ fs. We take the amplitude and duration of the electric field of the x-ray pulse to be such that the intensity of the x-ray pulse is $10^{13}$ W/cm$^2$ and $\tau_X = 0.5$ fs. We consider ionization from the $1\sigma$ (O side) and $2\sigma$ (N side) orbitals for the triplet state of NO$^+$. In our results, we fully account for the energy range of the x-ray pulse. That is, the Fourier transform of the x-ray pulse, since the full width at half maximum is equal to 0.5 fs, extends over energies roughly $\pm 4$ eV from the central photon energy. In Figs. 3(a) and 4(a), for photon energies of the x-ray pulse close to the ionization threshold, such as 544 eV (0.9 eV excess energy) for the $1\sigma$ orbital and 411 eV (0.8 eV excess energy) for the $2\sigma$ orbital, we show that we achieve control of the final angle of electron escape. That is, we achieve control when the electron due to the x-ray pulse is released at $t_{ion}$ in the IR pulse with very small momentum $k'(t_{ion}) = \sqrt{2(\hbar\omega - I_p)}$ compared to

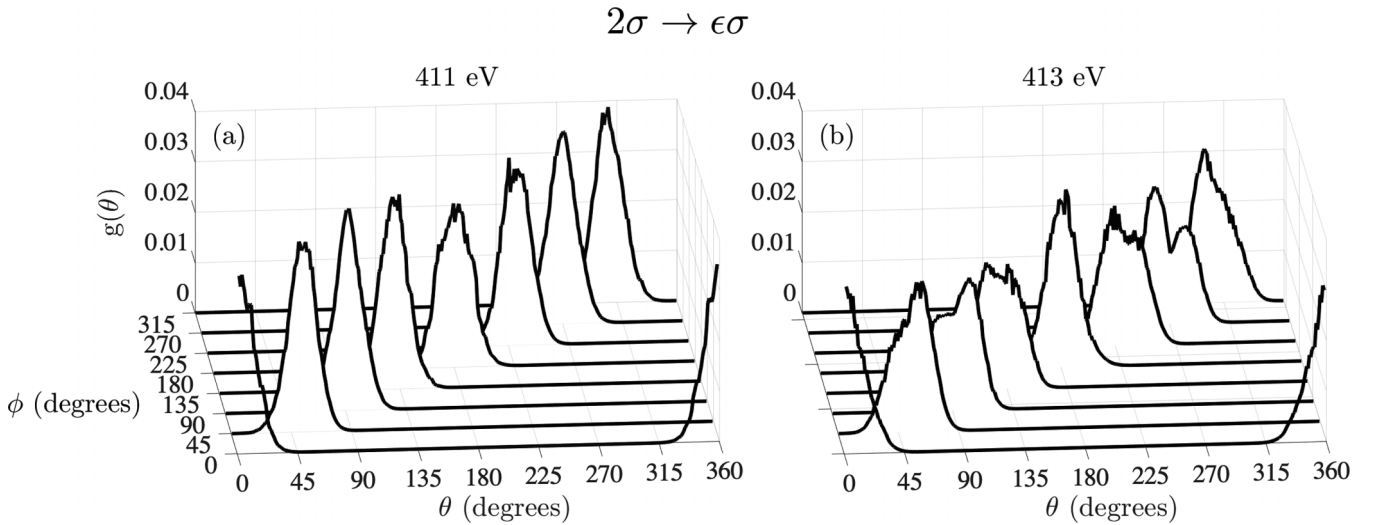

FIG. 4. Same as Fig. 3, but for an electron ionizing from the $2\sigma$ orbital of NO. The x-ray photon energy is (a) 411 eV and (b) 413 eV.

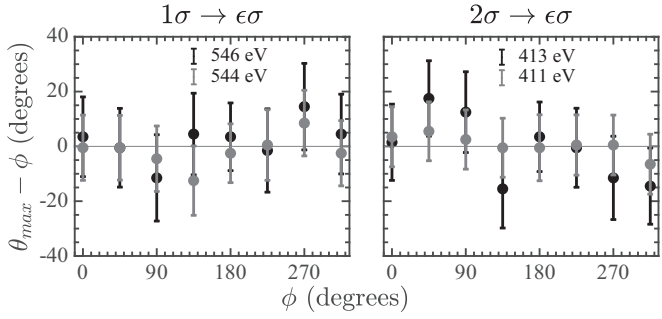

FIG. 5. The most probable angle of ejection, $\theta_{\max}$ (black dots), as a function of the delay $\phi$ between the x-ray and IR pulses, for an electron ionizing from the $1\sigma$ and $2\sigma$ orbitals. The intensity of the IR pulse is $5 \times 10^{13}$ W/cm$^2$. The vertical grey bars denote the standard deviation of the probability distribution g($\theta$).

the momentum the electron gains from the IR pulse (0.6 a.u.). For the 544 eV and 411 eV photon energies considered here the excess momentum of the ionized electron is 0.26 a.u and 0.24 a.u., respectively. Indeed, Figs. 3(a) and 4(a) show that for each $\phi$ the distribution g($\theta$) is narrow and centered around the angle $\theta_{\max}$ of $\theta$ that corresponds to the maximum g($\theta$), that is, $\theta_{\max} = \phi$. This means that there is a one-to-one mapping between the phase delay $\phi$ and the most probable angle of electron escape, $\theta_{\max}$. However, as we increase the photon energy of the x-ray pulse and hence the excess energy of the electron released due to the x-ray pulse at $t_{\text{ion}}$ in the IR pulse, Figs. 3(b) (2.9 eV excess energy) and 4(b) (2.8 eV excess energy) show that the distributions of g($\theta$) become more wide. Hence, for higher photon energies, we do not control the final angle of electron escape as well as for smaller photon energies. Quite interestingly, for high photon energies the distributions of g($\theta$) preserve features of the angular patterns of ionization in the presence of just the x-ray pulse. This is clearly seen for $\phi = 90°, 270°$ for a transition from the $1\sigma$ orbital when the x-ray energy is 543 eV [Fig. 3(b)] as well as for a transition from the $2\sigma$ orbital when the photon energy is 413 eV [Fig. 4(b)]. Indeed, in Fig. 3(b), for $\phi = 90°, 270°$, we find that g($\theta$) has a double-peak structure. The momentum gain of the ionizing electron from the IR pulse points along the $+x$ axis for $\phi = 90°$, while it points along the $-x$ axis for $\phi = 270°$. Hence, the higher probability of an electron to be ejected along the O side just in the presence of the x-ray pulse [see Fig. 2(a)] gives rise to the higher-valued peak at $\theta < 90°$ for $\phi = 90°$ and at $\theta > 90°$ for $\phi = 270°$. For the transition from the $2\sigma$ orbital in Fig. 4(b), we observe a reversed pattern of the higher-value peak of g($\theta$). This is due to an electron having a higher probability to be ejected along the N side just in the presence of the x-ray pulse [see Fig. 2(c)]. In addition, we better illustrate in Fig. 5 the one-to-one mapping between $\theta_{\max}$ and the phase delay $\phi$ between the x-ray and IR pulses by plotting $\theta_{\max} - \phi$ as a function of $\phi$. We do so for the transitions from the $1\sigma$ and $2\sigma$ orbitals for the same photon energies of the x-ray pulses as the ones considered in Figs. 3 and 4. For smaller photon energies of the x-ray pulse, we find that for each $\phi$ the values of $\theta_{\max} - \phi$ are close to zero; i.e., they lie close to the black horizontal line at zero in Fig. 5. Furthermore, for each $\phi$, the standard deviation of

$\theta$ with respect to the probability distribution g($\theta$) is smaller for the lower photon energies. This is seen by the shorter (smaller range in degrees) error bars for the smaller photon energies (grey bars) compared to the longer error bars for the larger photon energies (black bars). The difference between smaller and larger photon energies is especially evident for the transition from the $2\sigma$ orbital, since a 2-eV increase in the photon energy of the x-ray pulse is relatively larger for the $2\sigma$ orbital that has a lower threshold energy. The small spread of the angles $\theta$ around $\theta_{\max}$, for each $\phi$, means we achieve excellent control of electron motion. On a final note, in the framework of the strong-field approximation considered in this work after the electron is released in the IR pulse due to the x-ray pulse, it is evident that the electron motion is completely determined by the IR pulse if the electron is released in the IR pulse with zero excess energy. In this case the final momentum of the electron is completely determined by the vector potential at the time of ionization which forms an angle $\phi$ with the $z$ axis and thus the final angle of electron ejection, $\theta$, is equal to $\phi$. In this section we have demonstrated that we achieve control of electron motion with the IR pulse for a range of photon energies of the x-ray pulse above the threshold energy for a given transition.

### E. Streaking of electron ionization by an IR pulse

In the following, we obtain doubly differential final electron momenta distributions for low intensities of the circularly polarized IR pulse. We show that many of the features of these doubly differential electron distributions correspond to angular patterns of electron ionization just in the presence of the x-ray pulse. We plot the final electron momenta distributions on the $x-z$ plane of the IR pulse for an electron ionizing from the $1\sigma$ (Fig. 6) and the $2\sigma$ (Fig. 7) orbitals for the triplet state of NO$^+$. To obtain the probability for an electron to escape with final momentum components $p_{fx}$ and $p_{fz}$ regardless of the component $p_{fy}$, we perform the following integration:

$$|\mathcal{A}(p_{fx}, p_{fz})|^2 = \int dp_{fy} |\mathcal{A}(\vec{p}_f)|^2. \qquad (32)$$

In Figs. 6 and 7, we plot $|\mathcal{A}(p_{fx}, p_{fz})|^2$, which is the doubly differential probability for an electron to ionize on the $x-z$ plane with final momenta ($p_{fx}, p_{fz}$), for a low intensity of the circularly polarized IR pulse equal to $5 \times 10^{12}$ W/cm$^2$. The photon energy and FWHM of the IR pulse are $\omega_{\text{IR}} = 2300$ nm and $\tau_{\text{IR}} = 100$ fs. The amplitude and duration of the electric field of the x-ray pulse are such that the intensity of the x-ray pulse is equal to $10^{13}$ W/cm$^2$ and $\tau_X = 0.5$ fs. Here, unlike in Sec. III D, there is no phase delay between the x-ray and IR pulse; i.e., $\phi = 0$. We obtain these momenta distributions when the NO molecule is on the $x-z$ plane for various orientations $\theta_{\text{mol}}$ with respect to the $z$ axis. The angle $\theta_{\text{mol}}$ corresponds to the angle $\beta$ in Euler angle notation (see Sec. II A). We vary the angle $\theta_{\text{mol}}$ from 0 to $\pi$ in steps of $\frac{\pi}{4}$.

We take the photon energies of the x-ray pulse to be equal to 561 eV for the $1\sigma$ orbital (Fig. 6) and 428 eV for the $2\sigma$ orbital (Fig. 7), which is roughly 18 eV above their respective ionization thresholds. Hence, at the time $t_{\text{ion}}$, the electron ionizes due to the x-ray pulse with momentum roughly equal to 1.15 a.u. Since the x-ray pulse is taken to have a short duration, the times of ionization that we sample are close to the center of the IR pulse. Hence, from Eq. (9), the momentum gain of the ionizing electron from the IR pulse is equal to $-A_{\text{IR}}(t_{\text{ion}}) = -A_{\text{IR}}(\approx 0) = \frac{E_0^{\text{IR}}}{\omega_{\text{IR}}} = 0.60$ a.u. From Eq. (8), we thus find that the maximum and minimum final momenta of the electron are 1.75 and $-0.55$ a.u., respectively. Indeed, these maximum and minimum values of the momentum are seen in the right-hand columns of Figs. 6 and 7 for $p_{fx} \approx 0$.

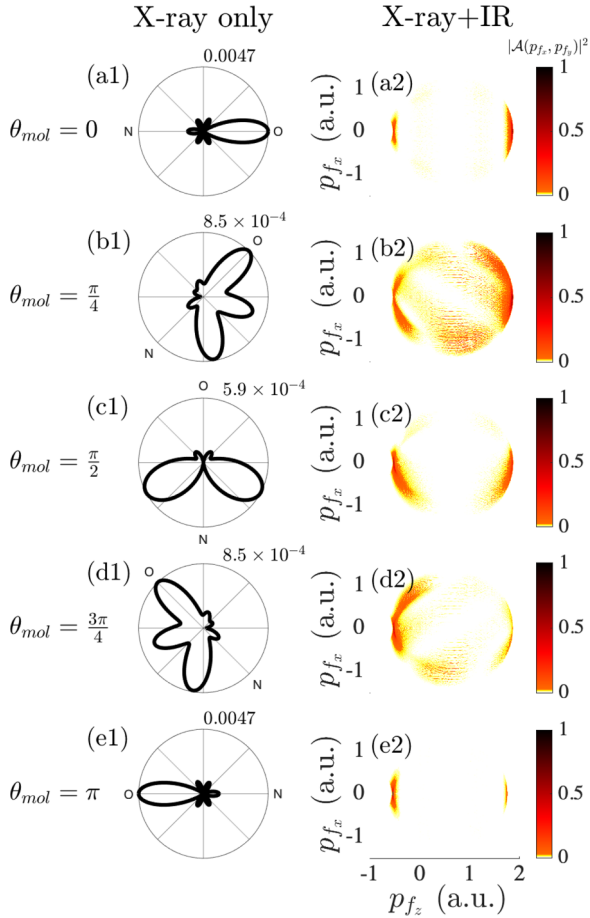

FIG. 6. For the triplet state of NO$^+$ for ionization from the $1\sigma$ orbital and a photon energy of the x-ray pulse equal to 561 eV, on the left column, we polar plot $|D_0(\vec{k}')|^2$ and project on the $x-z$ plane, i.e., we integrate over all angles $\phi_X$. On the right column, we plot the doubly differential probability $|\mathcal{A}(p_{fx}, p_{fz})|^2$ for an electron to escape on the plane of the circularly polarized IR pulse, which is the $x-z$ plane. The color plots are divided by their respective maximum differential probability to give the same relative scale for all color plots. The intensity of the circularly polarized IR pulse is equal to $5 \times 10^{12}$ W/cm$^2$. The phase delay between the x-ray and IR pulses, $\phi$, is equal to $0°$. The right and left column plots are obtained for the NO molecule being on the $x-z$ plane with an angle $\theta_{\text{mol}}$, measured with respect to the $z$ axis, ranging from 0 to $\pi$ in steps of $\frac{\pi}{4}$.

Next, we show that the momenta distributions obtained from the x-ray + IR pulses in the right-hand columns of Figs. 6 and 7 exhibit features consistent with the angular pattern of ionization solely due to the x-ray pulse. We show this by pairing each color plot of $|\mathcal{A}(p_{fx}, p_{fz})|^2$ (right-hand column of Figs. 6 and 7) with the respective polar plot of $|D_0(\vec{k}')|^2$ (left-hand column of Figs. 6 and 7). We note that in the polar plots, we plot the projection of $|D_0(\vec{k}')|^2$ on the $x-z$ plane, i.e., we integrate over the angle $\phi_X$, in order to better compare with the color plots of $|\mathcal{A}(p_{fx}, p_{fz})|^2$. Starting with the molecular orientation $\theta_{\text{mol}} = 0$, we observe that the high-probability lobe towards the O side in Fig. 6(b1) corresponds to the high-probability lobe in Fig. 6(a1). The smaller lobes in Fig. 6(a1) that are located at angles different than $0°$ and $180°$ are shifted compared to Fig. 6(b1) due to the momentum gain along the $z$ axis from the IR pulse. This is true for all angles $\theta_{\text{mol}}$ considered in Figs. 6 and 7. For the molecular orientations $\theta_{\text{mol}} = 0, \frac{\pi}{4}, \frac{3\pi}{4}, \pi$, we see that a significant probability in the color plots of Figs. 6 and 7 corresponds to a large probability for the electron to ionize along the O side for ionization from the $1\sigma$ orbital and along the N side for ionization from the $2\sigma$ orbital (see the corresponding large lobes in the polar plots). However, this is not the case for $\theta_{\text{mol}} = \frac{\pi}{2}$. In Figs. 6(b3) and 7(b3), we see that when the molecule is perpendicular to the x-ray pulse the electron does not escape along the molecular axis. In this latter case, the electron escapes with larger probability at angles roughly equal to $120°$ with respect to the O (N) side for ionization from the $1\sigma$ ($2\sigma$) orbital. These smaller probability lobes are present for all angles $\theta_{\text{mol}}$.

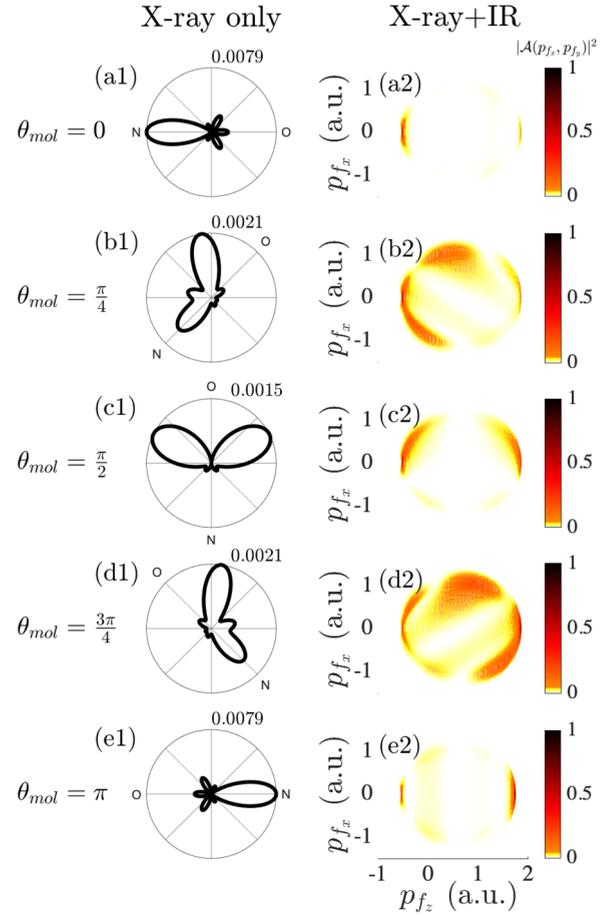

FIG. 7. Same as Fig. 6, but for ionization from the $2\sigma$ orbital with a photon energy of the x-ray pulse equal to 428 eV.

Overall, comparing the doubly differential plots of $|\mathcal{A}(p_{fx}, p_{fz})|^2$ with the polar plots of $|D_0(\vec{k}')|^2$ we find that, for small intensities of the IR pulse, most features in the former plots correspond to angular patterns of ionization in the presence of just the x-ray pulse. This is the reason why doubly differential plots of an electron to ionize with momenta $(p_{fx}, p_{fz})$ are used to extract photoionization time delays [26].

Finally, we note again that the Coulomb potential is fully accounted for the interaction of the NO molecule with the x-ray pulse. We neglect the Coulomb potential only during the propagation inside the IR pulse of the electron released at time $t_{\text{ion}}$. We expect that this approximation will not affect our findings concerning control and streaking of the electron motion. Fully accounting for the Coulomb potential at all stages will most probably result in broader distributions $g(\theta)$, for the high-intensity IR case, and doubly differential probability $|\mathcal{A}(p_{fx}, p_{fz})|^2$ for the low-intensity IR case.

## IV. CONCLUSION

In conclusion, we have shown how to obtain continuum molecular wavefunctions for open-shell molecules in the Hartree-Fock framework. We have obtained these wavefunctions when the total spin symmetry of the open-shell orbital and the orbital where the electron ionizes from is singlet or triplet. Using these continuum wavefunctions, we have obtained dipole matrix elements for ionization of an electron due to a linearly polarized x-ray pulse.

Following ionization from the x-ray pulse, we have investigated the effects on final electron escape when we streak with a circularly polarized infrared pulse. These effects depend on the intensity of the streaking infrared pulse. For a high intensity, we have shown that we control the angle of an electron escaping to the continuum. This control is achieved by varying the phase delay between the ionizing x-ray pulse and the infrared pulse. When the photon energy of the x-ray pulse is very close to the ionization threshold, a one-to-one mapping between the phase delay and the final angle of electron escape is achieved. For a low intensity, we have shown that the momentum distributions on the plane of the infrared pulse roughly image the angular electron escape patterns resulting by the x-ray pulse. The information obtained from the latter momentum distributions is of use in computing photoionization time delays.

## ACKNOWLEDGMENTS

The authors A.E. and M.E.M. acknowledge the use of the UCL Myriad High Throughput Computing Facility (Myriad@UCL), and associated support services, in the completion of this work. A.E. acknowledges the Leverhulme Trust Research Project Grant No. 2017-376. M.E.M. acknowledges funding from the EPSRC Project No. 2419551. This work was motivated by experimental capabilities at LCLS and supported on the SLAC side by the U.S. DOE, Office of Science, Office of Basic Energy Sciences (BES), Accelerator and Detector Research program, and the DOE-BES Chemical Sciences, Geosciences, and Biosciences Division (CSGB).

[1] M. Drescher, M. Hentschel, R. Kienberger, M. Uiberacker, V. Yakovlev, A. Scrinzi, Th. Westerwalbesloh, U. Kleineberg, U. Heinzmann, and F. Krausz, Time-resolved atomic inner-shell spectroscopy, Nature (London) **419**, 803 (2002).

[2] S. Biswas, B. Förg, L. Ortmann, J. Schötz, W. Schweinberger, T. Zimmermann, L. Pi, D. Baykusheva, H. A. Masood., I. Liontos, A. M. Kamal, N. G. Kling, A. F. Alharbi, M. Alharbi, A. M. Azzeer, G. Hartmann, H. J. Wörner, A. S. Landsman, and M. F. Kling, Probing molecular environment through photoemission delays, Nat. Phys. **16**, 778 (2020).

[3] A. L. Cavalieri, N. Müller, Th. Uphues, V. S. Yakovlev, A. Baltuška, B. Horvath, B. Schmidt, L. Blümel, R. Holzwarth, S. Hendel, M. Drescher, U. Kleineberg, P. M. Echenique, R. Kienberger, F. Krausz, and U. Heinzmann, Attosecond spectroscopy in condensed matter, Nature (London) **449**, 1029 (2007).

[4] W. B. Li, R. Montuoro, J. C. Houver, L. Journel, A. Haouas, M. Simon, R. R. Lucchese, and D. Dowek, Photoemission in the no molecular frame induced by soft-x-ray elliptically polarized light above the $N(1s)^{-1}$ and $O(1s)^{-1}$ ionization thresholds, Phys. Rev. A **75**, 052718 (2007).

[5] R. Pazourek, S. Nagele, and J. Burgdörfer, Attosecond chronoscopy of photoemission, Rev. Mod. Phys. **87**, 765 (2015).

[6] M. Mountney, G. P. Katsoulis, S. H. Møller, K. Jana, P. B. Corkum, and A. Emmanouilidou, Mapping the direction of electron ionization to phase delay between VUV and IR laser pulses, Phys. Rev. A **106**, 043106 (2022).

[7] W. Helml, I. Grguraš, P. N. Juranić, S. Düsterer, T. Mazza, A. R. Maier, N. Hartmann, M. Ilchen, G. Hartmann, L. Patthey, C. Callegari, J. T. Costello, M. Meyer, R. N. Coffee, A. L. Cavalieri, and R. Kienberger, Ultrashort free-electron laser x-ray pulses, Appl. Sci. **7**, 915 (2017).

[8] W. Helml, A. R. Maier, W. Schweinberger, I. Grguraš, P. Radcliffe, G. Doumy, C. Roedig, J. Gagnon, M. Messerschmidt, S. Schorb, C. Bostedt, F. Grüner, L. F. DiMauro, D. Cubaynes, J. D. Bozek, Th. Tschentscher, J. T. Costello, M. Meyer, R. Coffee, S. Düsterer *et al.*, Measuring the temporal structure of few-femtosecond free-electron laser x-ray pulses directly in the time domain, Nat. Photonics **8**, 950 (2014).

[9] I. Grguraš, A. R. Maier, C. Behrens, T. Mazza, T. J. Kelly, P. Radcliffe, S. Düsterer, A. K. Kazansky, N. M. Kabachnik, Th. Tschentscher, J. T. Costello, M. Meyer, M. C. Hoffmann, H. Schlarb, and A. L. Cavalieri, Ultrafast x-ray pulse characterization at free-electron lasers, Nat. Photonics **6**, 852 (2012).

[10] P. M. Maroju, M. Di Fraia, O. Plekan, M. Bonanomi, B. Merzuk, D. Busto, I. Makos, M. Schmoll, R. Shah, P. R. Ribič, L. Giannessi, G. De Ninno, C. Spezzani, G. Penco, A. Demidovich, M. Danailov, M. Coreno, M. Zangrando, A. Simoncig, M. Manfredda *et al.*, Attosecond coherent control of electronic wave packets in two-colour photoionization using a

novel timing tool for seeded free-electron laser, Nat. Photonics **17**, 200 (2023).

[11] N. Hartmann, G. Hartmann, R. Heider, M. S. Wagner, M. Ilchen, J. Buck, A. O. Lindahl, C. Benko, J. Grünert, J. Krzywinski, J. Liu, A. A. Lutman, A. Marinelli, T. Maxwell, A. A. Miahnahri, S. P. Moeller, M. Planas, J. Robinson, A. K. Kazansky, N. M. Kabachnik *et al.*, Attosecond time-energy structure of x-ray free-electron laser pulses, Nat. Photonics **12**, 215 (2018).

[12] S. Li, Z. Guo, R. N. Coffee, K. Hegazy, Z. Huang, A. Natan, T. Osipov, D. Ray, A. Marinelli, and J. P. Cryan, Characterizing isolated attosecond pulses with angular streaking, Opt. Express **26**, 4531 (2018).

[13] J. Duris, S. Li, T. Driver, E. G. Champenois, J. P. MacArthur, A. A. Lutman, Z. Zhang, P. Rosenberger, J. W. Aldrich, R. Coffee, G. Coslovich, F.-J. Decker, J. M. Glownia, G. Hartmann, W. Helml, A. Kamalov, J. Knurr, J. Krzywinski, M.-F. Lin, J. P. Marangos *et al.*, Tunable isolated attosecond x-ray pulses with gigawatt peak power from a free-electron laser, Nat. Photonics **14**, 30 (2020).

[14] S. Li, T. Driver, P. Rosenberger, E. G. Champenois, J. Duris, A. Al-Haddad, V. Averbukh, J. C. T. Barnard, N. Berrah, C. Bostedt, P. H. Bucksbaum, R. N. Coffee, L. F. DiMauro, L. Fang, D. Garratt, A. Gatton, Z. Guo, G. Hartmann, D. Haxton, W. Helml *et al.*, Attosecond coherent electron motion in Auger-Meitner decay, Science **375**, 285 (2022).

[15] D. C. Haynes, M. Wurzer, A. Schletter, A. Al-Haddad, C. Blaga, C. Bostedt, J. Bozek, H. Bromberger, M. Bucher, A. Camper, S. Carron, R. Coffee, J. T. Costello, L. F. DiMauro, Y. Ding, K. Ferguson, I. Grguraš, W. Helml, M. C. Hoffmann, M. Ilchen *et al.*, Clocking Auger electrons, Nat. Phys. **17**, 512 (2021).

[16] P. Eckle, M. Smolarski, P. Schlup, J. Biegert, A. Staudte, M. Schöffler, H. G. Muller, R. Dörner, and U. Keller, Attosecond angular streaking, Nat. Phys. **4**, 565 (2008).

[17] J. Itatani, F. Quéré, G. L. Yudin, M. Y. Ivanov, F. Krausz, and P. B. Corkum, Attosecond Streak Camera, Phys. Rev. Lett. **88**, 173903 (2002).

[18] H. I. B. Banks, D. A. Little, J. Tennyson, and A. Emmanouilidou, Interaction of molecular nitrogen with free-electron-laser radiation, Phys. Chem. Chem. Phys. **19**, 19794 (2017).

[19] S. K. Semenov, N. A. Cherepkov, G. H. Fecher, and G. Schönhense, Generalization of the atomic random-phase-approximation method for diatomic molecules: $N_2$ photoionization cross-section calculations, Phys. Rev. A **61**, 032704 (2000).

[20] J. W. Davenport, Ultraviolet Photoionization Cross Sections for $N_2$ and CO, Phys. Rev. Lett. **36**, 945 (1976).

[21] G. R. Wight, C. E. Brion, and M. J. Van Der Wiel, K-shell energy loss spectra of 2.5 keV electrons in $N_2$ and CO, J. Electron Spectrosc. Relat. Phenom. **1**, 457 (1972).

[22] M. E. Smith, V. McKoy, and R. R. Lucchese, Multipletspecific shape resonant features in photoionization of NO, J. Chem. Phys. **82**, 4147 (1985).

[23] F. Holzmeier, J. Joseph, J. C. Houver, M. Lebech, D. Dowek, and R. R. Lucchese, Influence of shape resonances on the angular dependence of molecular photoionization delays, Nat. Commun. **12**, 7343 (2021).

[24] M. Y. Ivanov and O. Smirnova, Ionization in strong low-frequency fields: From quantum *S*-matrix to classical pictures, Lecture Notes, CORINF Network (unpublished).

[25] A.-T. Le, H. Wei, C. Jin, and C. D. Lin, Strong-field approximation and its extension for high-order harmonic generation with mid-infrared lasers, J. Phys. B: At. Mol. Opt. Phys. **49**, 053001 (2016).

[26] M. Schultze, M. Fieß, N. Karpowicz, J. Gagnon, M. Korbman, M. Hofstetter, S. Neppl, A. L. Cavalieri, Y. Komninos, Th. Mercouris, C. A. Nicolaides, R. Pazourek, S. Nagele, J. Feist, J. Burgdörfer, A. M. Azzeer, R. Ernstorfer, R. Kienberger, U. Kleineberg, E. Goulielmakis *et al.*, Delay in photoemission, Science **328**, 1658 (2010).

[27] L. Gallmann, I. Jordan, H. J. Wörner, L. Castiglioni, M. Hengsberger, J. Osterwalder, C. A. Arrell, M. Chergui, E. Liberatore, U. Rothlisberger, and U. Keller, Photoemission and photoionization time delays and rates, Struct. Dyn. **4**, 061502 (2017).

[28] P. V. Demekhin, A. Ehresmann, and V. L. Sukhorukov, Single center method: A computational tool for ionization and electronic excitation studies of molecules, J. Chem. Phys. **134**, 024113 (2011).

[29] M. E. Rose, *Elementary Theory of Angular Momentum* (Wiley, New York, 1957).

[30] J. Pagaran, S. Fritzsche, and G. Gaigalas, Maple procedures for the coupling of angular momenta. IX. Wigner *D*-functions and rotation matrices, Comput. Phys. Commun. **174**, 616 (2006).

[31] H. Dachsel, Fast and accurate determination of the Wigner rotation matrices in the fast multipole method, J. Chem. Phys. **124**, 144115 (2006).

[32] R. Y. Rubinstein and D. P. Froese, *Simulation and the Monte Carlo Method*, 2nd ed. (Wiley, New York, 2007).

[33] B. H. Bransden and C. J. Joachain, *Physics of Atoms and Molecules* (Pearson Education India, 2003).

[34] H. Goldstein, *Classical Mechanics* (Addison-Wesley, Reading, MA, 1980).

[35] F. L. Pilar, *Elementary Quantum Chemistry* (Courier Corporation, 2001).

[36] H.-J. Werner and P. J. Knowles, A second order multiconfiguration SCF procedure with optimum convergence, J. Chem. Phys. **82**, 5053 (1985).

[37] H.-J. Werner, P. J. Knowles, G. Knizia, F. R. Manby, and M. Schütz, MOLPRO: A general-purpose quantum chemistry program package, WIRE: Comput. Mol. Sci. **2**, 242 (2012).

[38] T. H. Dunning, Gaussian basis sets for use in correlated molecular calculations. I. The atoms boron through neon and hydrogen, J. Chem. Phys. **90**, 1007 (1989).

[39] R. P. Orenha and S. E. Galembeck, Molecular orbitals of NO, $NO^+$ and $NO^-$: A computational quantum chemistry experiment, J. Chem. Educ. **91**, 1064 (2014).

[40] R. E. Stratmann, R. W. Zurales, and R. R. Lucchese, Multipletspecific multichannel electroncorrelation effects in the photoionization of NO, J. Chem. Phys. **104**, 8989 (1996).

[41] P. Lin and R. R. Lucchese, Theoretical studies of core excitation and ionization in molecular systems, J. Synchrotron Radiat. **8**, 150 (2001).

[42] J. Breidbach and L. S. Cederbaum, Universal Attosecond Response to the Removal of an Electron, Phys. Rev. Lett. **94**, 033901 (2005).

[43] K. Hosaka, J. Adachi, M. Takahashi, and A. Yagishita, N 1$s$ photoionization cross sections of nitric oxide molecules in the shape resonance region, J. Phys. B: At. Mol. Opt. Phys. **36**, 4617 (2003).

[44] C. E. Brion and K. H. Tan, Photoelectron branching ratios and partial oscillator strengths for the photoionization of NO (20–60 eV), J. Electron Spectrosc. Relat. Phenom. **23**, 1 (1981).

[45] S. Wallace, D. Dill, and J. L Dehmer, Shape resonant features in the photoionization spectra of NO, J. Chem. Phys. **76**, 1217 (1982).

[46] M. N. Piancastelli, The neverending story of shape resonances, J. Electron Spectrosc. Relat. Phenom. **100**, 167 (1999).

[47] J. J. García-Ripoll, P. Zoller, and J. I. Cirac, Speed Optimized Two-Qubit Gates with Laser Coherent Control Techniques for Ion Trap Quantum Computing, Phys. Rev. Lett. **91**, 157901 (2003).

[48] B. Scharfenberger, W. J. Munro, and K. Nemoto, Coherent control of an NV$^-$ center with one adjacent $^{13}$C, New J. Phys. **16**, 093043 (2014).

[49] J. I.-J. Wang, D. Rodan-Legrain, L. Bretheau, D. L. Campbell, B. Kannan, D. Kim, M. Kjaergaard, P. Krantz, G. O. Samach, F. Yan, J. L. Yoder, K. Watanabe, T. Taniguchi, T. P. Orlando, S. Gustavsson, P. Jarillo-Herrero, and W. D. Oliver, Coherent control of a hybrid superconducting circuit made with graphene-based van der Waals heterostructures, Nat. Nanotechnol. **14**, 120 (2019).

[50] P. B. Corkum and F. Krausz, Attosecond science, Nat. Phys. **3**, 381 (2007).

[51] W. Boutu, S. Haessler, H. Merdji, P. Breger, G. Waters., M. Stankiewicz, L. J. Frasinski, R. Taieb, J. Caillat, A. Maquet, P. Monchicourt, B. Carre, and P. Salieres, Coherent control of attosecond emission from aligned molecules, Nat. Phys. **4**, 545 (2008).

[52] A. Hache, J. E. Sipe, and H. M. van Driel, Quantum interference control of electrical currents in gas, IEEE J. Quantum Electron. **34**, 1144 (1998).

[53] M. Viteau, A. Chotia, M. Allegrini, N. Bouloufa, O. Dulieu, D. Comparat, and P. Pillet, Optical pumping and vibrational cooling of molecules, Science **321**, 232 (2008).

[54] C.-Y. Lien, C. M. Seck, Y.-W. Lin, J. H. V. Nguyen, D. A. Tabor, and B. C. Odom, Broadband optical cooling of molecular rotors from room temperature to the ground state, Nat. Commun. **5**, 4783 (2014).

[55] E. Dupont, P. B. Corkum, H. C. Liu, M. Buchanan, and Z. R. Wasilewski, Phase-Controlled Currents in Semiconductors, Phys. Rev. Lett. **74**, 3596 (1995).

[56] S. Sederberg, F. Kong, F. Hufnagel, C. Zhang, E. Karimi, and P. B. Corkum, Vectorized optoelectronic control and metrology in a semiconductor, Nat. Photonics **14**, 680 (2020).

[57] K. Jana, K. R. Herperger, F. Kong, Y. Mi, C. Zhang, P. B. Corkum, and S. Sederberg, Reconfigurable electronic circuits for magnetic fields controlled by structured light, Nat. Photonics **15**, 622 (2021).

[58] S. Sederberg, F. Kong, and P. B. Corkum, Tesla-Scale Terahertz Magnetic Impulses, Phys. Rev. X **10**, 011063 (2020).

[59] K. Jana, E. Okocha, S. H. Møller, Y. Mi, S. Sederberg, and P. B. Corkum, Reconfigurable terahertz metasurfaces coherently controlled by wavelength-scale-structured light, Nanophotonics **11**, 787 (2022).